\documentclass[aps,prd,showpacs,twocolumn,preprintnumbers,amsmath,amssymb,epsf,nofootinbib,tightenlines]{revtex4}

\usepackage{amsmath}
\usepackage{graphicx}
\usepackage{color}

\def\gtorder{\mathrel{\raise.3ex\hbox{$>$}\mkern-14mu
	\lower0.6ex\hbox{$\sim$}}}
\def\ltorder{\mathrel{\raise.3ex\hbox{$<$}\mkern-14mu
	\lower0.6ex\hbox{$\sim$}}}

\def\lsim{\mathrel{\rlap{\lower4pt\hbox{\hskip1pt$\sim$}}
    \raise1pt\hbox{$<$}}}
\def\gsim{\mathrel{\rlap{\lower4pt\hbox{\hskip1pt$\sim$}}
    \raise1pt\hbox{$>$}}}

\def\beqn{\begin{eqnarray}}
\def\eeqn{\end{eqnarray}}
\def\barr{\begin{array}}
\def\earr{\end{array}}
\def\btab{\begin{tabular}}
\def\etab{\end{tabular}}
\def\bite{\begin{itemize}}
\def\eite{\end{itemize}}
\def\bcen{\begin{center}}
\def\ecen{\end{center}}

\def\eq{\begin{equation}}
\def\ee{\end{equation}}
\def\nn{\nonumber}

\def \beqn{\begin{eqnarray}}
\def \eeqn{\end{eqnarray}}

\def \bea{\begin{eqnarray}}
\def \beq{\begin{equation}}
\def \eea{\end{eqnarray}}
\def \eeq{\end{equation}}
\def \nn{\nonumber}
\def \bwt{\begin{widetext}}
\def \ewt{\end{widetext}}

\begin{document}

\title{Two-photon exchange correction to $2S$-$2P$ splitting in muonic helium-3 ions}


\author{Carl E. Carlson}
\email{carlson@physics.wm.edu}
\affiliation{College of William and Mary, 
         Physics Department,
         Williamsburg, Virginia 23187, USA}
\author{Mikhail Gorchtein}
\email{gorshtey@kph.uni-mainz.de}
\author{Marc Vanderhaeghen}
\affiliation{Institut f\"ur Kernphysik, Johannes Gutenberg-Universit\"at,   Mainz, Germany 
and\\
PRISMA Cluster of Excellence,  Johannes Gutenberg-Universit\"at, Mainz, Germany}

\begin{abstract}
We calculate the two-photon exchange correction to the Lamb shift in muonic helium-3 ions within the dispersion relations framework.  Part of the effort entailed making analytic fits to the electron-$^3$He quasielastic scattering data set, for purposes of doing the dispersion integrals.  Our result is that the energy of the 2$S$ state is shifted downwards by two-photon exchange effects by 15.14(49) meV, in good accord with the result obtained from a potential model and effective field theory calculation.
\end{abstract}
\date{\today}

\maketitle

\section{Introduction}
Lamb shift measurements in muonic helium are underway to measure the nuclear radius of the helium isotopes \cite{Antognini:2011:Conf:PSAS2010}.  The motivation comes from the proton radius puzzle, where the reported proton radii from measurements involving electrons and measurements involving muons have been different, with the difference exceeding five standard deviations~\cite{Pohl:2010zza, Antognini:1900ns}.  For reviews, see~\cite{Pohl:2013yb,Carlson:2015jba}.  One can hope to learn more about the root cause of the discrepancy by seeing if it persists, and how large its effect may be, with nuclei heavier than the proton.  To this end, experiments have been performed to measure the $2S$-$2P$ Lamb shift energy splitting in muonic deuterium, $^3$He, and $^4$He~\cite{Pohl:2016glp,Antognini:2015moa}.

\begin{figure}[htbp]
\begin{center}
\includegraphics[height=2cm]{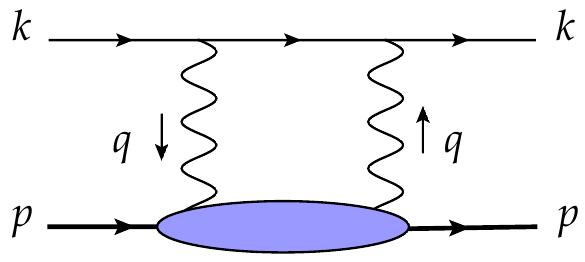}
\caption{The diagram showing the two-photon exchange correction in muonic $^3$He.}
\label{tpediag}
\end{center}
\end{figure}

The experiments obtain the radius from the deviation of the energy splitting measured from the energy splitting calculated for a pointlike nucleus.  To isolate the nuclear radius dependent term, it is crucial to know all the theory corrections that are large enough to affect the answer.  The Lamb shift $2S$-$2P$ energy splitting is given as
\begin{align}
\label{eq:basic}
\Delta E_{\rm Lamb} = \Delta E_{\rm QED} + \frac{ m_r^3 Z^4 \alpha^4}{12} R_E^2 
	+ \Delta E_{\rm TPE}		\,.
\end{align}
The accuracy of the QED term is not in question; reviews may be found in~\cite{Borie:2012zz,Antognini:2013jkc}. The second term will yield the charge radius~\cite{Karplus:1952zza,Eides:2000xc}.  The reduced mass $m_r$ is the usual 
\begin{align}
m_r = \frac{ m M_T }{ m + M_T }		,
\end{align}
where $m$ is the mass of the lepton and $M_T$ is the mass of the nucleus.  The third term is the two-photon exchange (TPE) correction, the subject of this note for the case of muonic $^3$He, and given diagrammatically in Fig.~\ref{tpediag}.

\smallskip

An important question to consider at the outset is how accurately the two-photon correction needs to be calculated.  An answer can be obtained from the uncertainty of the $\mathcal O(\alpha^4)$ charge radius term as predicted using the charge radius measured in electron scattering.  An analysis of world data for electron scattering on $^3$He~\cite{Sick:2014yha} quoted $R_E(^3{\rm He}) = 1.973(14)$ fm.  One can obtain a slightly better uncertainty limit by using the more extensive and more precise $^4$He electron scattering data together with isotope shift measurements from atomic spectroscopy.  The charge radius of $^4$He is
\begin{align}
R_E(^4{\rm He}) = 1.681(4) {\rm\ fm}		.
\end{align}
The isotope shift measures $\delta R_E^2 = R_E^2(^3{\rm He}) - R_E^2(^4{\rm He})$.  Unfortunately, the three existing measurements are not in agreement,
\beqn
\delta R_E^2= \left\{
	\begin{array}{ll}
	1.066(4) {\rm\ fm}^2  &  {\rm\ Ref.}~[13]
	\\
	1.074(4) {\rm\ fm}^2  &  {\rm\ Refs.}~[14,15]
	\\
	1.028(11) {\rm\ fm}^2  &  {\rm\ Ref.}~[16]
	\end{array}
\right.
\eeqn
These yield $R_E(^3{\rm He}) = 1.9728(36)$, $1.9748(36)$, and $1.9631(44)$ fm, respectively, adding uncertainties in quadrature.  
To exclude either of the above determinations by three respective standard deviations via a Lamb shift measurement, the second term in Eq. (\ref{eq:basic}) should be determined with an accuracy of about $1.5$ meV. Hence, the requirement to the precision of the third term of that equation, the TPE correction is to be below $1.5$ meV.  Its size, as we shall see, is about $15$ meV,  so as a fraction  one needs an accuracy better than $10\%$. One should bear in mind that even when this precision goal is met, the accuracy of the TPE calculation will remain by far the main limitation of the charge radius extraction, since the experimental accuracy of order $0.07$ meV or better \cite{Antognini:2011:Conf:PSAS2010} is expected. 


An additional numerical benchmark follows from what may happen if beyond the standard model (BSM) explanations of the proton radius are correct~\cite{TuckerSmith:2010ra,Batell:2011qq, Carlson:2012pc}.  In this scenario, the muonic $2S$-$2P$ energy deficit that was attributed to a smaller proton radius is instead attributed to a muon specific BSM force.  For purposes of benchmarking, consider a BSM model where the new exchange particle couples on the hadron side in proportion to the electric charge, like a dark photon that is muon specific on the lepton side (for an alternative scenario where its couplings to the proton and the neutron allowed arbitrary values, see Ref. \cite{Liu:2016qwd}).  Also consider, at least at the outset, that the new force is short range for both $\mu$-H and $\mu$-$^3$He.  This requires that the new exchange particle is heavy enough, and a few $10$'s of MeV will suffice.  Then the $330\,\mu$eV energy deficit for muonic hydrogen scales to
\begin{align}
\Delta E_{\rm BSM}^{\mu\,^3{\rm He}} =  6.0 {\rm\ meV}	,
\end{align}
for the $2S$-$2P$ splitting.  The bulk of the scaling comes from a $Z^4$ factor and the remainder from differences in the reduced mass.  Thus also from considering the scale of possible BSM effects, a $5$-$10\%$ calculation of the TPE correction is useful and relevant.  (A lower mass BSM exchange particle will reduce the value obtained for $\Delta E_{\rm BSM}^{\mu\,^3{\rm He}}$.)

An accurate potential model calculation of the TPE is already available \cite{Dinur:2015vzv,Hernandez:2016jlh}, so one may ask why another estimate is useful?  The answer is that the result is very important for the study of the proton radius puzzle, so that another calculation using a very different technique is worth doing and reporting.  Our fully relativistic calculation is directly phenomenological, using dispersion theory to connect electron-$^3$He elastic and inelastic scattering data to quantities that enter the evaluation of the TPE effect.  The already available calculation is nonrelativistic with relativistic corrections and is based on nuclear potential models.  The potentials are either a classical one, the AV18 potential  abetted with three-nucleon forces, or a chiral effective field theory potential, also with three-nucleon forces added to the two-nucleon ones.  We will see that the dispersive and the nuclear potential model calculations corroborate each other.

Dispersive evaluations of the TPE correction have been carried out for muonic hydrogen~\cite{Pachucki:1999zza,Martynenko:2005rc,Carlson:2011zd,Birse:2012eb,Gorchtein:2013yga} and muonic deuterium~\cite{Carlson:2013xea}.  For $\mu$-H they represent the state of the art and are accepted as such~\cite{Antognini:1900ns}.  Other methods evaluating TPE in $\mu$-H~\cite{Nevado:2007dd,Alarcon:2013cba,Peset:2015zga} are not yet equivalent in accuracy.  The deuterium situation is different.  The deuteron is loosely bound, easily polarized, and can be broken up with just a bit over $2$ MeV energy transfer.  The relevant integrals for the dispersive evaluation are weighted toward low-energy transfer and low-momentum transfer.  Electron-deuteron scattering data is currently sparse in these regions, and the outcome is a not very stringent $35\%$ uncertainty in the dispersive result~\cite{Carlson:2013xea}.  One must rely instead on nuclear potential model evaluations~\cite{Pachucki:2011xr,Friar:2013rha,Hernandez:2014pwa,Pachucki:2015uga} (we refer the reader to a summary of theoretical calculations in Ref. \cite{Krauth:2015nja}).  Helium nuclei are tightly bound compared to the deuteron, and more than $5$ MeV energy transfer is required for $^3$He disintegration.  It is enough to make a significant difference.  There are more data points than for the deuteron in the range where the necessary integrals have their main support, and the higher threshold for the low energy weighting makes the numerical results smaller.  We find that for $^3$He we can meet the accuracy goal.

\section{Calculation}

The diagram that contains the nuclear and hadronic structure-dependent
${\cal O}(\alpha^5)$ correction to the Lamb shift is shown in
Fig. \ref{tpediag}.  The lower part of the diagram, the blob containing
nuclear and hadronic structure dependence is encoded in the forward
virtual Compton tensor,
\beqn
&&T^{\mu\nu}=\frac{i}{8\pi M_T}\int d^4x	\,	e^{iqx}\langle
p|T\,j^\mu(x)j^\nu(0)|p\rangle\\
&&=\left(-g^{\mu\nu}+\frac{q^\mu
    q^\nu}{q^2}\right)T_1(\nu,Q^2)
+\frac{\hat p^\mu\hat p^\nu}{M_T^2} T_2(\nu,Q^2),\nn
\eeqn
where $\hat p^\mu=p^{\mu}-\frac{p\cdot q}{q^2}q^\mu$, $Q^2=-q^2$, 
$\nu=(p\cdot q)/M_T$ and $M_T$ is the $^3$He mass.  A target spin average is implied.  
Following \cite{Carlson:2011zd}, we can write the contribution of the
two-photon exchange diagram to the $n\ell$ energy level as 
\beqn
&&\Delta E_{n\ell}=\frac{8\alpha^2m}{\pi}\phi^2_{n\ell}(0)\int d^4Q\\
&&\times\frac{(Q^2+2Q_0^2)T_1(iQ_0,Q^2)-(Q^2-Q_0^2)T_2(iQ_0,Q^2)}{Q^4(Q^4+4m^2Q_0^2)},\nn
\eeqn
where a Wick rotation $q_0=iQ_0$ was made, and
$\phi^2_{n\ell}(0)=\mu_r^3(Z\alpha)^3/(\pi n^3)\delta_{\ell0}$. The amplitudes $T_{1,2}(\nu,Q^2)$ are even functions of $\nu$ and
their imaginary parts are related to the spin-independent structure
functions of lepton-$^3$He scattering,
\beqn
{\rm Im}T_1(\nu,Q^2)&=&\frac{1}{4M_T}F_1(\nu,Q^2)	,	\nn\\
{\rm Im}T_2(\nu,Q^2)&=&\frac{1}{4\nu}F_2(\nu,Q^2).
\eeqn

Before writing the dispersion relation, we will give the Born terms, which are obtained from the elastic box and crossed box version of Fig.~\ref{tpediag} and $^3$He electromagnetic vertex,
\beqn
\Gamma^\mu(q)=F_D(Q^2)\gamma^\mu+F_P(Q^2)i\sigma^{\mu\alpha}\frac{q_\alpha}{2M_T} ,
\eeqn
with $F_{D,P}$ the Dirac and Pauli form factors of $^3$He and $q$  the momentum of an incoming photon. 
To disambiguate, we will use $F_{D,P}^{p,n}$ notation for the proton and neutron Dirac and Pauli form factors, respectively.
The Born terms are
\begin{align}
T_1^B(q_0,Q^2) &= \frac{Z^2}{4\pi M_T} 	
\left\{ 	\frac{Q^4 G_M^2(Q^2) }
		{(Q^2-i\epsilon)^2 - 4M_T^2 q_0^2} - F_D^2(Q^2) 
	\right\}		\,,
					\nonumber\\
T_2^B(q_0,Q^2) &= \frac{Z^2M_T  Q^2}{ \pi (1+\tau_T) }  
	\frac{G_E^2(Q^2) + \tau_T G_M^2(Q^2)}
		{(Q^2-i\epsilon)^2 - 4M_T^2 q_0^2}	\,.
\end{align}
Nuclear electric and magnetic Sachs form factors are defined in the standard way,
\beqn
G_E&=&F_D-\tau_T F_P,\nn\\
G_M&=&F_D+F_P,
\eeqn
and $\tau_T=Q^2/(4M_T^2)$. 
The Born terms are useful for correctly obtaining the imaginary parts of the nucleon pole terms, but not reliable in general, since the given vertex assumes the incoming and outgoing nucleons are both on shell.  

We also define
\begin{align}
\overline T_{1,2}(\nu,Q^2) = T_{1,2}(\nu,Q^2) - T_{1,2}^{\rm pole}(\nu,Q^2)	,
\end{align}
where $T_{1,2}^{\rm pole}$ are the pole parts of the Born amplitudes.  For future use, the non-pole
$\overline T_1(0,Q^2)$ amplitude can be written as a term visible in the Born term plus a term proportional to $Q^2$ at small $Q^2$,
\begin{align}
\label{eq:magpol}
\overline T_1(0,Q^2) = - \frac{Z^2F_D^2(Q^2)}{ 4\pi M_T }
	+ \frac{ Q^2 }{ e^2 } \beta_M^{^3{\rm He}}(Q^2)		\,,
\end{align}
where $\beta_M^{^3{\rm He}}(0) = \beta_M^{^3{\rm He}}$ is the magnetic polarizability of $^3$He.

Given the known high-energy behavior of the structure functions, the
two amplitudes obey the following form of dispersion relation, 
\begin{align}
{\rm Re\,} \overline T_1(q_0,Q^2)&= \overline T_1(0,Q^2)
+\frac{q_0^2}{2\pi M_T}\int\limits_{\nu_{\rm th}}^\infty\frac{d\nu F_1(\nu,Q^2)}{\nu(\nu^2-q_0^2)},
				\nn\\
{\rm Re\,} \overline T_2(q_0,Q^2)&=
\frac{1}{2\pi}\int\limits_{\nu_{\rm th}}^\infty\frac{d\nu F_2(\nu,Q^2)}{\nu^2-q_0^2}	,
\end{align}
where the integrals are evaluated in the principle value sense, and $\nu_{\rm th}$ is the inelastic threshold.

We divide the contribution to the energy shift of the $S$-state into three physically distinct terms that originate 
from the subtraction term $\overline T_1(0,Q^2)$, the nucleon pole, and finally all
excited intermediate states that may couple to $\gamma N$, respectively 
\beqn
\Delta E_{nS}&=& \Delta E_{nS}^{\rm subt} + \Delta E_{nS}^{\rm el}+ \Delta E_{nS}^{\rm inel}.  \label{split} 
\eeqn
with 
\begin{align}						\label{eq:subt}
\Delta E_{nS}^{\rm subt} &= \frac{4\pi\alpha^2}{m}\phi^2_{n0}(0)	\nn\\
&	\times
	\int_0^\infty\frac{dQ^2}{Q^2}\frac{\gamma_1(\tau_l)}{\sqrt\tau_l} 
	\left[ \overline T_1(0,Q^2) - \overline T_1(0,0)  \right] ,
\end{align}
\begin{align}
&\Delta E_{nS}^{\rm el} =  \alpha^2 Z^2\phi^2_{n0}(0)
\int_0^\infty\frac{dQ^2}{Q^2}	\Bigg\{	  \frac{ 16 m M_T }{ (M_T+m) Q } 
	G'_E(0)
\label{lamb} 
													\nn\\
&	- \frac{ m }{ M_T(M_T^2-m^2) }
	\Bigg[ \!\!	\left(\frac{\gamma_1(\tau_T)}{\sqrt\tau_T}
-\frac{\gamma_1(\tau_l)}{\sqrt\tau_l}\right) \!\! \left( G_M^2 - 1 \right)	\nn\\
&	\qquad -\left(\frac{\gamma_2(\tau_T)}{\sqrt\tau_T}
-\frac{\gamma_2(\tau_l)}{\sqrt\tau_l}\right)		
	\frac{ G_E^2 - 1 + 
	\tau_T  \left( G_M^2 - 1 \right) }{\tau_T(1+\tau_T)}		\Bigg]
		\Bigg\}, 
\end{align}
\begin{align}
& \Delta E_{nS}^{\rm inel}=-\frac{2\alpha^2}{m M_T}\phi^2_{n0}(0)\int_0^\infty\frac{dQ^2}{Q^2}
\int_{\nu_{\rm th}}^\infty\frac{d\nu}{\nu}		\nn\\
&\times\left[\tilde\gamma_1(\tau,\tau_l)F_1(\nu,Q^2)
+\frac{M_T \nu}{Q^2}\tilde\gamma_2(\tau,\tau_l)F_2(\nu,Q^2)\right].\label{inel}
\end{align}
We introduced $\tau_l=Q^2/(4m_\ell^2)$, $\tau=\nu^2/Q^2$, and the auxiliary functions,
\beqn
\gamma_1(\tau)&\equiv&(1-2\tau)\sqrt{1+\tau}+2\tau^{3/2}\nn\\
\gamma_2(\tau)&\equiv&(1+\tau)^{3/2}-\tau^{3/2}-\frac{3}{2}\sqrt\tau\nn\\
\tilde\gamma_1(\tau,\tau_l)&\equiv&\frac{\sqrt\tau_l\gamma_1(\tau_l)-\sqrt\tau\gamma_1(\tau)}{\tau_l-\tau}\nn\\
\tilde\gamma_2(\tau,\tau_l)&\equiv&\frac{1}{\tau_l-\tau}\left(
\frac{\gamma_2(\tau)}{\sqrt\tau}-\frac{\gamma_2(\tau_l)}{\sqrt\tau_l}\right).
\eeqn
Furthermore, we have subtracted two-photon exchange terms in $\Delta E^{\rm el}$ that are already included in a bound state calculation.  The ``$-1$''s come from iterations within the basic wave equation calculation that gives the bound state, which is done for a pointlike nucleus, and the 
$G'_E$ term removes the iteration of the lowest order nuclear radius term seen in Eq.~(\ref{eq:basic}).  Recall that by definition,
\begin{align}
R_E^2   = -6 G'_E(0)	.
\end{align}

\subsection{Elastic contribution}
Using the form factor parametrization obtained by Amroun \textit{et al.}~\cite{Amroun:1994qj} and Sick \cite{Sick:2008zza} in the sum-of-gaussians form we obtain:
\beqn
\Delta E_{2S}^{el}&=&-10.93\,{\rm meV}
,
\eeqn
in an excellent agreement with a dedicated extraction of the Zemach radius in Ref. \cite{Sick:2014yha} from scattering data, which leads to the energy shift 
\beqn
\Delta E_{2S}^{el}&=&-10.87(27)\,{\rm meV},
\eeqn
where we note a significant $\sim$3\% uncertainty, and we will use it as an uncertainty estimate for our evaluation.

Krutov \textit{et al.}~\cite{Krutov:2015pxa} used an exponential form factor 
$G_E=\exp[-R_E^2Q^2/6]$, leading to an estimate of the elastic contribution,
\beqn
\Delta E_{2S}^{el}&=&-10.28\,{\rm meV},
\eeqn
considerably smaller than what one obtains by using phenomenological form factors
which fit the data better.  Similarly, the elastic contribution obtained using the deuteron form factors of~\cite{DeJager:1987qc} is also notably smaller than the value we quote above, but~\cite{DeJager:1987qc} did not have available the later data obtained by Amroun \textit{et al.}~\cite{Amroun:1994qj}.

\subsection{Inelastic contribution}

We separate the inelastic contributions into two regions, the quasieleastic or nuclear region, where the final states are either three nucleons or a deuteron plus a  proton, and the pion production or nucleon region.  In practice, we separate these regions at the pion production threshold.   Thus, we write the inelastic contributions as two parts, 
\beqn
\Delta E^{\rm inel}_{nS}=\Delta E^{\rm nuclear}_{nS}+\Delta E^{\rm nucleon}_{nS},
\eeqn 
which we will treat in the next two subsections. 

\subsubsection{Nuclear polarizability contribution}

The bulk of the data in this region has been tabulated by Benhar, Day, and Sick~\cite{Benhar:2006er,Benhar:2006wy}.  This tabulation has 83 data sets categorized by incoming electron energy and electron scattering angle.  To this tabulation, we add the three $180^\circ$ data sets~\cite{Jones:1979zza,Chertok:1969zz}.  

For purposes of evaluating the integrals, we make analytic fits to this data.  Details are given in the Appendix.  In brief, we started with functional forms motivated by a Fermi gas model of the nucleus, which was called to our attention by superscaling studies of electron-nucleus scattering~\cite{Maieron:2001it,Bosted:2012qc,Bodek:2014pka}.  We modified the forms with additional parameters so that we could fit the lower energy and lower momentum transfer data crucial to the present calculation.  We paid special attention to the photoproduction ($Q^2=0$) and near photoproduction data, and added extra terms to ensure these regions were well represented.

\begin{figure*}[t]
\begin{center}
\includegraphics[width=17cm]{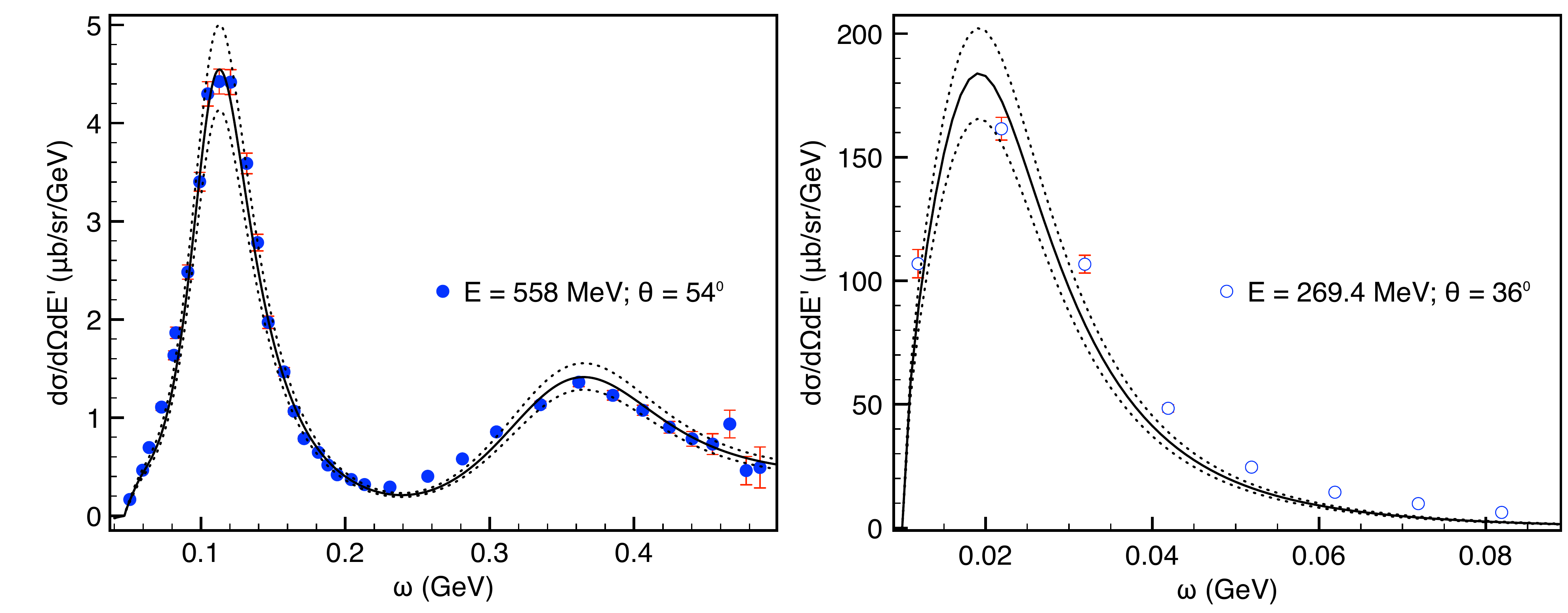}
\includegraphics[width=17cm]{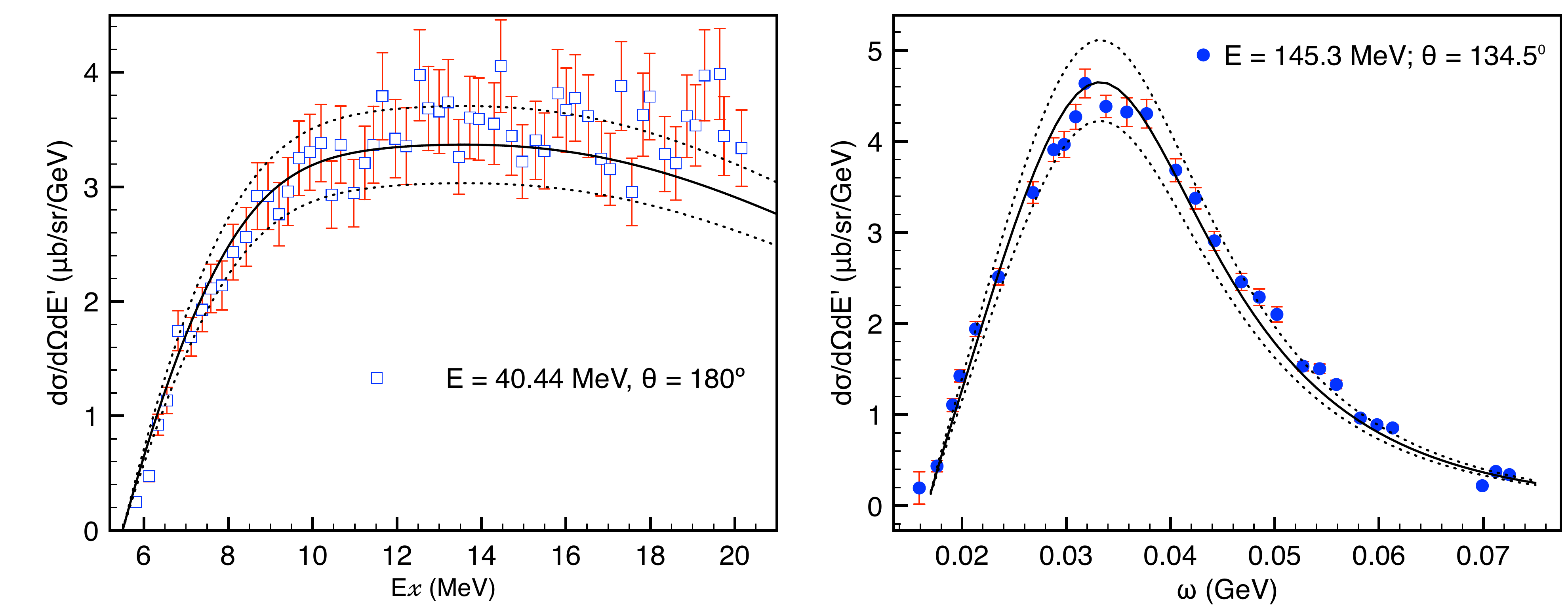}
\caption{Parametrization defined in Eqs. (\ref{eq:f12QE}-\ref{eq:fitreal2}) compared to the experimental data in forward (upper panels) and backward (lower panels) kinematics. Data are from Ref.~\cite{Jones:1979zza} (empty squares), Ref.~\cite{Marchand:1985us} (solid circles), and Ref.~\cite{Dow:1988rk} (empty circles). Dotted curves in all three panels indicate the 10\% uncertainty band. The spectrum in the lower left panel is shown as function of excitation energy $E_x$ in units of MeV, and as function of photon energy in units of GeV otherwise.}
\label{figDATA}
\end{center}
\end{figure*}

Samples of the fits are shown in Fig.~\ref{figDATA}, with $\pm 10\%$ error bands indicated.  At the $\pm 10\%$ level, the fits are overall good, and if the uncertainties in both the data and the fits are purely statistical, the uncertainty in the integrals is much less than $10\%$.

The result for the quasielastic or nuclear part of the inelastic contributions is
\begin{align}
\Delta E^{\rm nuclear} = - 5.50\,(40) {\rm\ meV}.
\end{align}
\indent
The uncertainty on this number is explained in appendix.

\subsubsection{Intrinsic nucleon polarizability contribution}

The contributions from the nucleon region, where we have energy sufficient for nucleon breakup, is separable from other contributions, and the results of this subsection can be taken and combined with calculations where the other contributions are calculated in ways different from what we have done.

We use modern helium-3 virtual photoabsorption data that were parametrized in terms
of resonances plus non-resonant background \cite{Bosted:2012qc,Bodek:2014pka} that capitalizes on the free proton and neutron fits of Refs. \cite{Bosted:2007xd, Christy:2007ve} with a Fermi-smearing effect built in. Since the integration over the energy extends beyond the validity of the fit of Refs. \cite{Bosted:2007xd, Christy:2007ve}, we supplement the correct high-energy behavior by adopting a Regge-behaved background, specified in our previous work for the deuteron case, \cite{Carlson:2013xea}, and adjusted for the case of the helium target. The result of the integration is
\beqn
\Delta E_{2S}^{\rm nucleon}&=&-0.306(15)\, {\rm meV}.
\eeqn

Summing the nuclear and nucleon polarizability contributions leads to
\begin{align}
\Delta E_{2S}^{\rm inel} = - 5.81 (40) {\rm\ meV},
\end{align}
with the uncertainties added in quadrature.\\

\subsection{Subtraction contribution}

The subtraction function $ T_1(0,Q^2)$ is generally unknown.  We need it at nonnegative $Q^2$.  Excepting $Q^2 = 0$, this is unphysical kinematics and not directly related to scattering data.  
Instead we obtain it from a sum rule~\cite{Gorchtein:2015eoa} based on the dispersion relation for $T_1$ and several additional observations.

The sum rule is fully explained in~\cite{Gorchtein:2015eoa}.  The dispersion relation for $T_1$ is analyzed and approximated after observing that: firstly, the imaginary part Im$\,T_1(\nu,Q^2) \propto F_1(\nu,Q^2)$ in the quasielastic region becomes quite small beyond a certain value of $\nu$ for a given $Q^2$, creating a ``gap region'' between it and the onset of the pion production region.  Secondly, there is a separation of scales, in that the gap region is quite broad.  Thirdly,  for high energies, the binding effects are relatively small and we can, at least within integrals, treat the $^3$He structure functions as just that for two protons plus one neutron.  Then combing the dispersion relation for $^3$He with the corresponding dispersion relations for the proton and neutron, one can obtain
\beqn
&&\overline T_1(0,Q^2)	=	\frac{1}{2\pi M_T}
	\int\limits_{\nu_{min}(Q^2)}^{\nu_{max}(Q^2)}
\frac{d\nu}{\nu}	F_1(\nu,Q^2)	\nn\\
&&\;\;\;+\frac{Q^2}{e^2}(2\beta_M^p(Q^2)+\beta_M^n(Q^2))
	- \frac{2 F_D^{p\,2} + F_D^{n\,2}}{4\pi M_p}	\label{eq:SubFunc}	.
\eeqn
Here, $\nu_{\rm max}(Q^2)$ is the upper limit of the region where the quasielastic structure function is large, and $\beta_M^{p,n}(Q^2)$ are the nucleon analogs of $\beta_M^{^3{\rm He}}(Q^2)$ described earlier, normalized at $Q^2 = 0$ to the experimentally determined magnetic polarizabilities of the proton and neutron, respectively.

A test of the sum rule is to evaluate its $Q^2 = 0$ limit, using the known result for $\overline T_1(0,0)$ to obtain,
\begin{equation}
NZ = 2 = 2 \int_{\nu_{min}}^{\nu_{max}}	\frac{d\nu}{\nu}	F_1(\nu,0)		\,,
\end{equation}
which is the Bethe-Levinger sum rule~\cite{Levinger:1950zz}.  Using the structure function fits described in the Appendix and integrating to $30$ MeV above threshold gives
\begin{equation}
2 \int_{\nu_{min}}^{\nu_{min} + 30 {\rm MeV}}	\frac{d\nu}{\nu}	F_1(\nu,0) = 1.65 \,.
\end{equation}
Integrating to $60$ MeV above threshold gives $1.78$.  The sum rule appears to work at the $15\%$ level or better.

Using the sum rule to obtain $T_1(0,Q^2)$ and $T_1(0,0)$ for the subtraction term energy, Eq.~(\ref{eq:subt}), leads to 
\beqn
\Delta E_{2S}^{\rm subt} = \left( 1.39 + 0.21 \right){\rm\, meV} = 1.60 {\rm\ meV}	,
\eeqn
where the separated terms are for the nuclear and nucleon contributions, respectively.   We took the nucleon polarizabilities from the PDG average~\cite{Agashe:2014kda},
 \begin{align}
\beta_M^p=2.5(0.4)\cdot10^{-4}\,{\rm fm}^3,&&\beta_M^n=3.7(2.0)\cdot10^{-4}\,{\rm fm}^3 ,
 \end{align}
 and took the $Q^2$ dependence for $\beta_M^{p,n}(Q^2)$ following~\cite{Birse:2012eb}.

The uncertainty in the subtraction term contribution to the Lamb shift comes from two sources which we refer to as statistical and systematic. 
The statistical one is due to the finite precision and kinematical coverage of the data used for evaluating the sum rule integral. This uncertainty should be considered jointly with the uncertainty of the inelastic contribution because the same parametrization of the data enters there. An important effect is a partial cancellation of the subtraction term and the inelastic $F_1$ contribution, which leads to a reduced uncertainty. We address this uncertainty in detail in the appendix. 

The systematic uncertainty is due to the use of the approximate sum rule for the subtraction function. To assess its uncertainty, we note that the derivation of the sum rule~\cite{Gorchtein:2015eoa} relies on the assumption of the large gap between the nuclear and nucleon excitation spectra, and of the dominance of the nuclear contributions over the hadronic ones. A comparison of the nuclear and nucleonic contributions to Eq. (\ref{eq:SubFunc}) reveals that they become of similar size starting from $Q^2 \approx (0.3$ GeV)$^2$. 
We assign a conservative 100\% uncertainty to the contribution coming from $Q^2$ beyond this value, and find
\beqn
(\Delta E_{2S}^{\rm subt, nuclear})_{Q^2>0.1\,{\rm GeV}^2}  =  0.11\,{\rm meV}	,
\eeqn
which leads us to the final estimate of the subtraction term,
\beqn
\Delta E_{2S}^{\rm subt} = 1.60(12)\,{\rm meV}.
\eeqn
where we added the uncertainty of the nucleon and nuclear contributions in quadrature.

Of palpable interest is the numerical value of the $^3$He polarizability $\beta_M^{^3{\rm He}}$, which can be obtained from the derivative of the sum rule at $Q^2 = 0$ and the relation in Eq.~(\ref{eq:magpol}).
This leads to
\begin{align}
\beta_M^{^3{\rm He}} &= \frac{ 2\alpha }{ M_T } 	\int_{\nu_0}^{\nu_{max}} \frac{ d\nu }{ \nu }
	\frac{ d F_1^{^3{\rm He}} }{ dQ^2 }(\nu, 0)
	- \frac{ Z^2 \alpha }{ 3 M_T } \big( R_E^{^3{\rm He}} \big)^2	\nn\\
&	+ \frac{ \alpha }{ 3M_p } \left[ 2R_{Ep}^2 + R_{En}^2 \right]
	+ 2 \beta_M^p + \beta_M^n	\,.
\end{align}
Using the $F_1$ parametrizations from the Appendix, the $^3$He charge radius 1.956 fm, the known nucleon radii, and the values for $\beta_M^{p,n}$ from Ref. \cite{Hagelstein:2015egb}, we obtain
\beqn
\beta_M^{^3{\rm He}} = 5.7 \times 10^{-3}\,{\rm fm}^3.
\eeqn
Since the value of the $^3$He magnetic polarizability is unknown, we stress that this is a prediction to be tested in the future, either experimentally or in an EFT calculation. Using the spread in the values of $\beta_M^{p,n}$ from different analyses and the uncertainty in the $Q^2$-slope of $F_1$ at low $Q^2$, we conjoin an uncertainty estimate,
\beqn
\beta_M^{^3{\rm He}}  = 5.7(0.5)\times10^{-3}\,{\rm fm}^3.
\eeqn

In contrast, a recent lattice calculation~\cite{Chang:2015qxa} gives a much smaller value 
$\beta^{^3{\rm He}}_M=5.4(2.2)\times10^{-4}\,{\rm fm}^3$, obtained in conjunction with a pion mass of about $806$ MeV. The same reference also suggests 
$\beta^d_M=4.4(1.5)\times10^{-4}\,{\rm fm}^3$, more than two orders of magnitude smaller than the EFT prediction of about $0.07$ fm$^3$~\cite{Friar:1997tr,Chen:1998vi}. 

In view of this inconclusive situation with the value of $\beta_M$, we wish to emphasize that as far as the nuclear polarizability contribution to the Lamb shift is concerned, it is practically insensitive to the  value of $\beta_M$. The reason for that is the cancellation between the inelastic and subtraction contributions, both coming from the transverse Compton amplitude $T_1$. The sum rule that we use here ensures that whenever the parametrization of the transverse QE data changes, this change is also propagated in the subtraction function, so that the net effect is small. This is in accord with the general expectation of smallness of the magnetic polarizability effect on Lamb shift, e.g. in potential model calculations.

\section{Results and discussion}

The result of our phenomenological, dispersion relation based, calculation is summarized in Table~\ref{tab}. 
The overall result is that the 2$S$ state has its energy shifted due to two-photon exchange by an amount
\begin{align}
\Delta E_{2S} = - 15.14 (49)  {\rm\, meV}.
\end{align}
The uncertainty limit is small enough, exceeding the criterion set out in the introduction by a factor of 3.

\begin{table}[t]
\begin{tabular}{l|r|r}
%
Contribution \ &  This work  & Refs. [21,22]
\\
\hline
\hline
 Elastic & $-10.93(27)$  & $-10.49(24)$ \\
 \qquad $\delta_{\rm Zem}^N$ &  &  $-0.52(3)$ \\
\hline
 Inelastic  & $-5.81(40)$	  & $-4.45(21)$ \\
 	\qquad Nuclear & $-5.50(40)$ &     $-4.17(17)$\\
 	\qquad Nucleon &  $-0.31(2)$ &  $-0.28(12)$\\
\hline
 Subtraction & $1.60(12)$   &\\
  	\qquad Nuclear & $1.39(12)$&	\\
 	\qquad Nucleon & $0.21(3)$& 	\\
 \hline
 \hline
Total TPE & $-15.14(49)$  &  $-15.46(39)$\\
 \hline
\end{tabular}
\caption{Individual contributions to $\Delta E_{2S}$ from two-photon exchange in $\mu$-$^3$He, in units of meV.}
\label{tab}
\end{table}

 To facilitate the comparison with other work, we included in Table~\ref{tab} results from the recent work of Ref. \cite{Dinur:2015vzv}.   We shall make some comparison, even though that calculation is very different from ours, so that comparisons of any but the total may be inexact.  
There, the nuclear elastic contribution,
\beqn
\Delta E_{2S}^{el}&=&-10.52(24)\,{\rm meV}, 
\eeqn
 is somewhat lower than our full elastic contribution, but once the nucleon Zemach correction $\delta_{\rm Zem}^N$ listed in the rightmost column in Table~\ref{tab} is added to it, their full result,
\beqn
\Delta E_{2S}^{el,\,tot.}&=&-11.01(25)\,{\rm meV},
\eeqn
is close to ours.

The same reference calculated the nuclear polarizability contribution using potential models and effective field theory. For a reasonable comparison, we should confront the sum of nuclear, nucleon polarizabilities, with the sum of the total inelastic and total subtraction contributions obtained in this work,
\beqn
-4.45(21)\;{\rm Refs.\ }[21,22]
&{\rm vs.}&-4.21(40)\;{\rm This \;work},\nn\\
\eeqn
where various uncertainties were added in quadrature.
The two results closely agree. 

We conclude by noting that the main limitation of our calculation of the TPE effect on the Lamb shift in muonic He-3 is the availability and precision of the quasielastic data at low $Q^2$ and forward angles. To assess the improvability of our result, we study the impact of possible measurements of the inclusive differential cross section for electron-$^3$He scattering in the quasielastic regime with the new MESA accelerator in Mainz. We assume a generic 5\% accuracy of the data, feasible for MESA with the laboratory energy $E=110$ MeV and scattering angle $20^\circ\leq\theta\leq30^\circ$, and list the projected accuracy in determining the parameter $a_2$ entering the parameterization of Eq.~(\ref{eq:qefix}), as well as that of the nuclear polarizability and the full TPE contributions to the Lamb shift in $\mu\,^3$He atoms in Table \ref{tab2}.  At the moment, the lowest available momentum transfer at forward angles is $Q^2=0.0091$ GeV$^2$ from the 110 MeV, $54^\circ$ data set of Ref. \cite{Dow:1988rk}, shown in Table \ref{tab2}.  We notice from Table II that a future 5~\% measurement of $d \sigma / d \Omega$ at MESA around $\theta \approx 30^\circ (20^\circ)$ will reduce the total uncertainty of the polarizability contribution by a factor of $2$ ($4$) respectively. The resulting TPE contribution will then be mainly limited by the present knowledge of the elastic contribution, which can also be improved by such future measurements. 

\begin{table}[]
\begin{tabular}{l|c|c|c}
%
Kinematics \ &  $\delta a_2$ \   & $\delta (\Delta E_{\rm2S}^{\rm nuclear})$ & $\delta (\Delta E_{\rm2S}^{\rm TPE})$\\
\hline
\hline
$E=110$ MeV  \ & & \\
\qquad $\theta=54^\circ$  & $\pm0.014$	  & $0.40$ meV & $0.49$ meV \\
\qquad $\theta=30^\circ$  & $\pm0.0075$	  & $0.21$ meV    & $0.35$ meV\\
 	\qquad $\theta=25^\circ$ & $\pm0.0055$ &  $0.16$ meV   & $0.33$ meV\\
 	\qquad $\theta=20^\circ$ &  $\pm0.0040$ & $0.11$ meV   & $0.30$ meV\\
 \hline
\end{tabular}
\caption{The impact of possible 5\% measurements of $d\sigma/d\Omega$ in the quasielastic kinematics at MESA with laboratory energy $E=110$ MeV and scattering angle $20^\circ\leq\theta\leq30^\circ$ for constraining the parameter $a_2$ and the nuclear polarizability and full TPE contributions to the Lamb shift in muonic $^3$He. The upper row shows the current situation with the available data.}
\label{tab2}
\end{table}

\acknowledgements{The authors are grateful to A. Antognini, J. Krauth and R. Pohl for valuable comments to the manuscript. CEC thanks the National Science Foundation (USA) for support under grant PHY-1516509.  MG and MV thank for support in the Deutsche Forschungsgemeinschaft DFG through the Collaborative Research Center CRC 1044.  MV thanks the College of William and Mary for its hospitality during the completion of this work.}

\appendix
\section{Fitting the quasielastic data}			\label{sec:aa}

We make use of the quasi elastic data collected on the Donal Day's web archive, which may be found through~\cite{Benhar:2006er}. The specific data that have the biggest impact on this calculation are \cite{Retzlaff:1994zz,Jones:1979zza,Marchand:1985us,Dow:1988rk} which extend from the quasielastic threshold to above the pion production threshold, and for 0.005 GeV$^2\leq Q^2\leq0.7$ GeV$^2$.

The full nuclear structure functions are parameterized here in two parts,
\begin{align}
F_{1,2}(\nu,Q^2) = F_{1,2}^{\rm QE}(\nu,Q^2) + F_{1,2}^{\rm real}(\nu,Q^2)	\,.
\end{align}

We use the super scaling parametrization of quasi elastic data according to \cite{Maieron:2001it,Bodek:2014pka,Bosted:2012qc} with additional adjustments to provide a better description of the data with an emphasis on the low energy and low $Q^2$ region. For the single-nucleon structure functions we use the following representation, 
\begin{align}
F_1^{\rm QE}&=
\frac{Q^2}{ | \vec q\, | }\bar G_M^2 S_1(\nu,Q^2)\label{eq:f12QE}	\nn\\
F_2^{\rm QE}&= \frac{\nu Q^4}{M_T | \vec q\, |^5 }\frac{\bar G_E^2+\tau_p
   \bar G_M^2}{1+\tau_p}( 2M_p + \nu )^2	S_2(\nu,Q^2),
\end{align}
where $\bar G_{E,M}^2=2(G_{E,M}^p)^2+(G_{E,M}^n)^2$, and the functions $S_i(\nu,Q^2)$ are defined as
\begin{align}
S_i(\nu,Q^2)=F(\psi') F_{P}(|\vec q\,|)\!
\left[1-\left(\frac{\nu_{thr}}{\nu}\right)^3\right]^{\alpha_i}\!\!\!
	\sqrt{\frac{\nu_{thr}}{\nu}}f_i^{QE}(Q^2),
\end{align}
which contain the superscaling variable $\psi'$, the super scaling function $F(\psi')$, and the Pauli suppression factor $F_{P}$, all as described in \cite{Bodek:2014pka}. Above, 
$\vec q\,^2=\nu^2+Q^2$,
$\tau_p = {Q^2}/(4m_p^2)$, 
$\nu_{\rm thr}=\epsilon_T+Q^2/(2M_T)$, with 
\begin{align}
\epsilon_T &= S_p + \frac{S_p^2}{2M_T}
	= M_d +M_p -M_T + \frac{S_p^2}{2M_T} 	\nn\\
	&= 5.493 {\rm\, MeV}	\,,
\end{align}
the helium-3 breakup threshold.
The parameters $\alpha_i$ are $\alpha_1=1$ and $\alpha_2=1/2$. The functions $f_i^{\rm QE}$ are obtained from a fit to the QE data in the vicinity of the QE peak, in the form
\beqn			\label{eq:qefix}
f_1^{\rm QE}(Q^2)&=&\frac{a_1+b_1\, Q^2}{1+b_1 \,Q^2},\\
f_2^{\rm QE}(Q^2)&=&\frac{a_2+b_2\,Q^2}{1+b_2 \,Q^2}f_1^{QE}(Q^2).\nn
\eeqn
\indent
The fit returned values $a_1=0.31(6)$, $b_1=54.4(7.0)$ GeV$^{-2}$, $a_2=0.014(14)$, $b_2=52(2)$ GeV$^{-2}$, and the numbers in the parentheses indicate the uncertainty. 

\begin{figure}
\includegraphics[width=8cm]{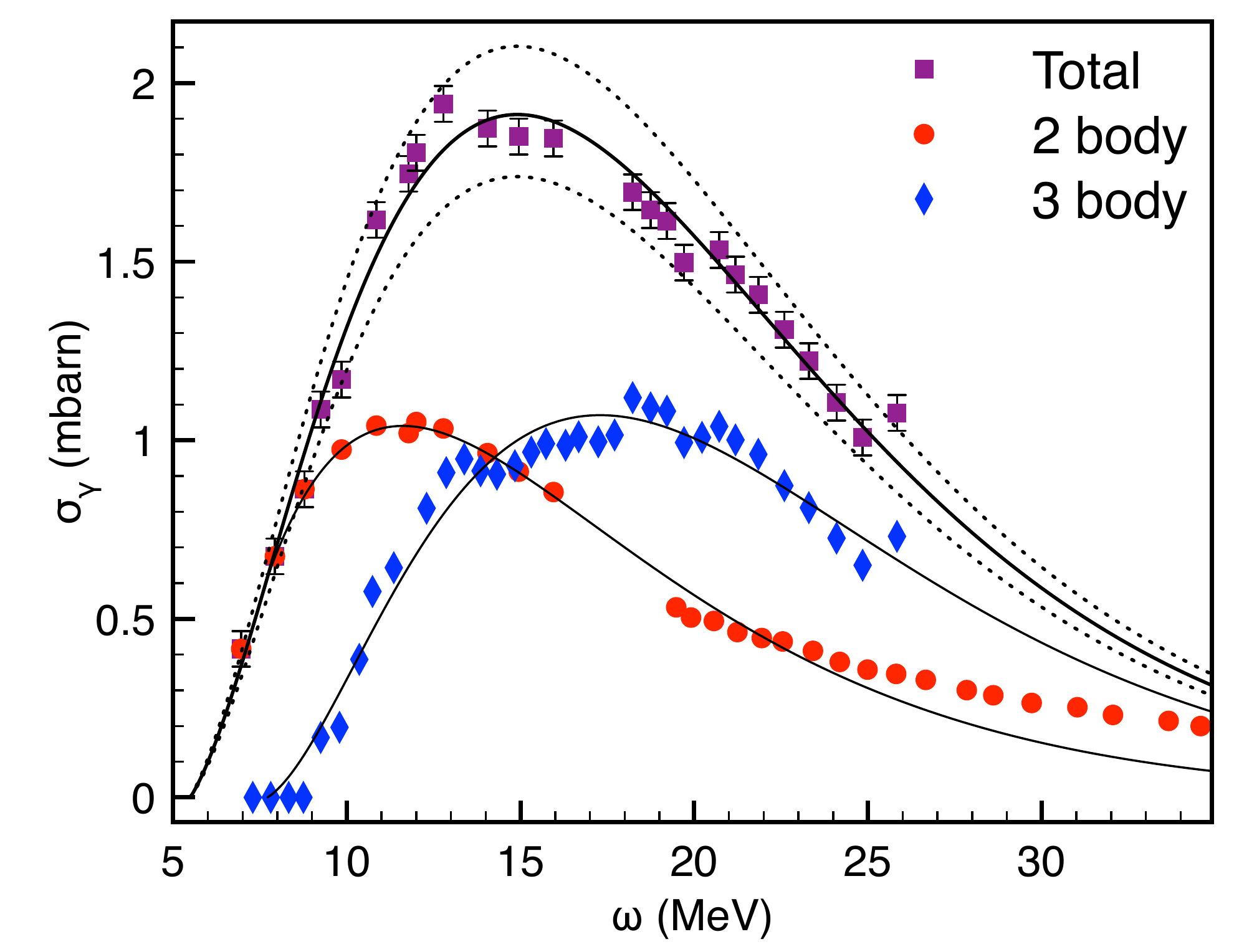}
\caption{Parametrization of the $^3$He photodisintegration $(p,d)$ data from Refs. \cite{Ticcioni:1973srh,Tornow:2011zz} (red circles) and 3-body data from Ref. \cite{Faul:1981zz} (blue rombs) are shown along with their sum (magenta squares) and the respective fits defined in Eqs. (\ref{eq:fitreal1},\ref{eq:fitreal2}).}
\label{figDATAreal}
\end{figure}

The structure functions defined above vanish at the real photon point, so we need to supplement a description at and near the real photoabsorption. This is done by fitting the real photon data first, and then extending this fit to finite values of $Q^2$. The real photon fit was done in the functional form
\beqn
\sigma_\gamma^{\rm tot}(\omega)=e^{-A(\omega-\epsilon_T)}\left[B(\omega-\epsilon_T)+C(\omega-\epsilon_T)^2\right],
\label{eq:fitreal1}
\eeqn
with the values of the parameters 
\beqn
A&=&0.200257\,{\rm MeV}^{-1},\nn\\ 
B&=&0.153202\,{\rm mb\,/\,MeV},\nn\\
C&=&0.125848\,{\rm mb\,/\,MeV}^2,\label{eq:fitreal2}
\eeqn
and is shown with the available data from Refs. \cite{Faul:1981zz,Ticcioni:1973srh,Tornow:2011zz} in Fig. \ref{figDATA}.  Note that the most recent 2-body data of Ref. \cite{Tornow:2011zz} exceeds the older data from \cite{Ticcioni:1973srh} at low energies. In Fig. \ref{figDATAreal} we combine the two data sets in one.

The structure functions are obtained according to 
\beqn
F_1^{\rm real}(\nu,Q^2)&=&\frac{M\omega}{4\pi^2\alpha}
	\sigma_\gamma^{\rm tot}(\omega)f_1^{\rm real}(Q^2)	\nn\\
F_2^{\rm real}(\nu,Q^2)&=&\frac{\nu Q^2}{M\vec q\,^2}F_1^{\rm real}(\nu,Q^2).
\eeqn
A relation between $F_1$ and $F_2$ of this form is equivalent to a vanishing of the respective contribution to the longitudinal cross section. Above,  
$\omega=\nu-Q^2/2M$,  and the function of $Q^2$ obtained from a fit to QE data is
\beqn
f_1^{\rm real}(Q^2)&=&\frac{1+a_3\,Q^2}{(1+b_3\,Q^4)^{c_3}},\label{eq:f1real}
\eeqn
with $a_3=167(6)$ GeV$^{-2}$, $b_3=94(16)$ GeV$^{-4}$ and $c_3=2.5(2)$.

The uncertainties of the parameters were obtained in the following way. At the first step, we fixed $f_2^{QE}=f_1^{QE}$ and $f_2^{real}=f_1^{real}$, and fitted 180$^\circ$ data by Jones et al. \cite{Jones:1979zza}, 144.5$^\circ$ data by Marchand et al. \cite{Marchand:1985us}, 134.5$^\circ$ data by Dow et al. \cite{Dow:1988rk}, as well as the transverse part of the L-T separated data by Retzlaff et al. \cite{Retzlaff:1994zz}. Because our fits were designed, in the first place, to provide a reliable input to the Lamb shift calculation where the integrals are weighted heavily towards low values of $Q^2$, we aimed at ensuring that we describe the data at lowest $Q^2$ values in the best possible way. In particular, the real photon data and the low-energy 180$^\circ$ data by Jones et al. \cite{Jones:1979zza} at $Q^2\approx 0.005$ GeV$^2$ and slightly above, fix the parameters $a_1$ and $a_3$, while they are not sensitive to other parameters. Fixing $a_{1,3}$ from a low-energy fit, we determined other parameters including other backward data. After the transverse part was determined, we turned to forward data at 36$^\circ$ and 60$^\circ$ by Marchand et al. \cite{Marchand:1985us} and 54$^\circ$ data by Dow et al. \cite{Dow:1988rk}, as well as data on the longitudinal response function by Retzlaff et al. \cite{Retzlaff:1994zz}. While no further modification was necessary for $f_2^{real}$, an additional adjustment of $f_2^{QE}$ at low values of $Q^2$ was required. The fit via the function $f_2^{QE}$ is, however, only determined for $Q^2\geq Q_{min}^2\approx0.009$ GeV$^2$, no data below that value are available at forward angles.  

The behavior of  $f_2^{QE}$ at lower virtualities, governed by the parameter $a_2$, is crucial for evaluating the Lamb shift. We proceeded as follows. Setting $a_2=0$ first, we obtained the reference value of the parameter $b_2=54.0(2.0)$ GeV$^{-2}$ from a fit at $Q^2\geq 0.02$ GeV$^2$, the value chosen to lie above $Q_{min}^2$. As the second step, we studied the extrapolation of $f_2^{QE}(a_2=0)$ to $0\leq Q^2\leq Q_{min}^2$ by means of a 3$^{\rm rd}$ order polynomial,
\beqn
g(Q^2,Q_{min}^2)=\sum_{n=0}^3 \frac{(Q^2-Q_{min}^2)^n}{n!}g_n.
\eeqn
We fixed its value and first two derivatives at $Q^2=Q^2_{min}$, i.e. $g_{0,1,2}$ to those of the function $f_2^{QE}(a_2=0,Q^2=Q_{min}^2)$ and treated $g_3$ as a free parameter with constraints:
\beqn
&&|g_3|\leq3\left|\left[f_2^{QE}(a_2=0,Q^2=Q_{min}^2)\right]'''\right|,\nn\\
&&g(Q^2,Q_{min}^2)\geq0\;\;\;{\rm for}\;\;\;Q^2\geq0,
\eeqn
the latter constraint being due to the fact that $F_2^{QE}$ is a cross section that is positive definite. This gives us the upper and lower value of $g(Q^2,Q_{min}^2)$ evaluated at $Q^2=0$ which we now identify with the parameter $a_2$. We also studied the dependence of choosing the matching point 0.009 GeV$^2\leq Q_{min}^2\geq0.017$ GeV$^2$, and obtained after averaging $a_2=0.014\pm0.014$. The uncertainty is a combined systematical and statistical one, but it is dominated by the systematical uncertainty, the one due to the extrapolation. Statistical uncertainty obtained from that of the parameter $b_2$ which fixes $\left[f_2^{QE}(a_2=0,Q^2=Q_{min}^2)\right]'''$ only contributes a couple percent. 

Finally, we used parameter $a_2$ as fixed, and refit of $f_2^{QE}$, allowing once again the parameter $b_2$ to vary. The resulting value $b_2=(52.0\pm2.0)$ GeV$^{-2}$ nicely agrees with the previously obtained $b_2=(54.0\pm2.0)$ GeV$^{-2}$, which serves as an a posteriori test of validity of this procedure. 

\begin{figure}
\includegraphics[width=9cm]{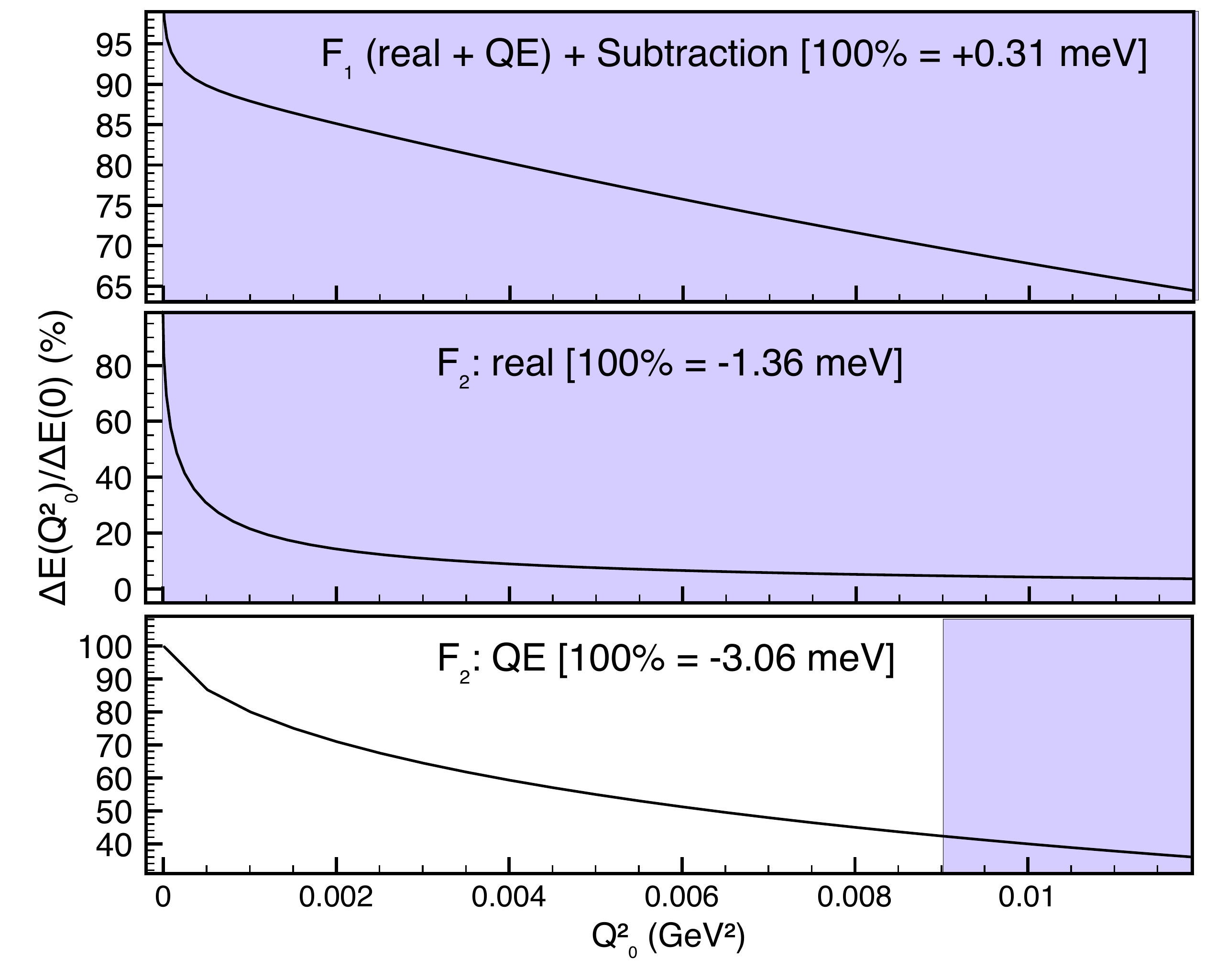}
\caption{Saturation of the dispersion integrals for  $\Delta E^{\rm subt} + \Delta E^{\rm inel} [F_1^{\rm QE} + F_1^{\rm real}]$ contribution (upper panel), $\Delta E^{\rm inel} [F_2^{\rm real}]$ (middle panel), $\Delta E^{\rm. inel} [F_2^{\rm QE}]$ (lower panel) as function of the lower limit of integration $Q_0^2$, in percent of the full result corresponding to $Q_0^2=0$. In each panel, the absolute value of each contribution to the Lamb shift in $\mu\,^3$He in meV is indicated. Shaded areas indicate the regions where experimental data are available.}
\label{figSaturation}
\end{figure}
To address the respective uncertainty for the Lamb shift calculation, we study the saturation of the dispersion integrals in Eqs. (\ref{eq:subt}, \ref{inel}), as function of the lower limit of the $Q^2$-integral, while the integral over energy is carried out over the full allowed range in the quasielastic region. In doing so, we consider the sum of the subtraction term and the inelastic contribution due to $F_1$, and separately contributions of $F_2^{\rm real}$ and $F_2^{\rm QE}$ to Eq. (\ref{inel}) as function of the lower limit of integration $Q_0^2$ in the range $0\leq Q_0^2\leq Q_{\rm min}^2$and take the value of each of these integrals with $Q_0^2=0$ for 100\%. Results are shown in Fig. \ref{figSaturation}. 
We observe that the sum $\Delta E^{\rm subt}+\Delta E^{{\rm inel},\,F_1}$ is not very sensitive to the details of the extrapolation below $Q_{\rm min}^2\approx0.005$ GeV$^2$: only about 20\% of the total of that contribution to the Lamb shift comes from $Q_0^2\leq Q_{\rm min}^2$, which in absolute values is mere 0.06 meV. 

Instead, both $F_2$ contributions are very sensitive to the lower limit of the integration: for $F_2^{\rm real}$ about 94\% comes from $Q_0^2\leq Q_{\rm min}^2$, whereas for $F_2^{\rm QE}$ about 40\% comes from that range. We recall that the former contribution is fixed at the extremes of the explored range, by real photon data at $Q^2=0$ and by the QE data at $Q^2=Q_{\rm min}^2$. We conclude that the uncertainty of $\Delta E^{\rm inel,\,real}$ is given by the uncertainties of the data. However, this is not the case for the QE contribution that is only fixed at $Q^2\geq Q_{\rm min}^2$. The fit to data above that value is consistent with the function $f_2^{QE}$ vanishing at $Q^2=0$, but it going to a finite small positive number instead is also not excluded. We expect the extrapolating function to be smooth because the point $Q_{\rm min}^2\approx0.009$ GeV$^2$ corresponds roughly to a distance of 2 fm, on the exterior of the charge distribution of $^3$He that has a radius of 1.97 fm, so we expect no structure beyond this point. 

It is seen that already a 5\% measurement at largest angle in the considered range will reduce the uncertainty of the nuclear polarizability contribution by a factor of $\sim2$, leaving the Zemach contribution the main source of the uncertainty in the full TPE calculation. Further improvement in the precision of the inelastic contribution will only have a marginal effect, unless a more precise determination of the 3$^{\rm rd}$ Zemach moment from elastic $e-\,^3$He scattering data will become possible. 

\bibliography{bib_TPE_2016}{}

\end{document}